
%
%
%
%
\magnification=1200
\hsize=13.5cm

\def\sk{\vskip .4cm}
\def\noi{\noindent}

\def\unmezzo{{1 \over 2}}
\def\epsi{\varepsilon}
\def\we{\wedge}

\def\de{\delta}

\def\part{\partial}

\def\R#1#2{ R^{#1}_{~~#2} }

\def\C#1#2{ {\bf C}_{#1}^{~~~#2} }

\def\q#1{   {{q^{#1} - q^{-#1}} \over {q^{\unmezzo}-q^{-\unmezzo}}}  }

\def\qm{q^{-1}}

\def\q1{$q \rightarrow 1$}
\def\Fmn{F_{\mu\nu}}
\def\Am{A_{\mu}}
\def\An{A_{\nu}}
\def\dm{\part_{\mu}}
\def\dn{\part_{\nu}}
\def\Ana{A_{\nu]}}
\def\Bna{B_{\nu]}}
\def\Zna{Z_{\nu]}}
\def\dma{\part_{[\mu}}
\def\qsu{$[SU(2) \times U(1)]_q~$}

\def\gij{g_{ij}}
\def\L{{\cal L}}

\def\gu{g_{U(1)}}
\def\gsu{g_{SU(2)}}
\def\tg{ {\rm tg} }

\hskip 10cm \vbox{\hbox{DFTT-19/92}\hbox
{April 1992}}
\vskip 0.6cm
\centerline{\bf  GAUGE THEORIES OF QUANTUM GROUPS}
\vskip 3cm
\centerline{\bf Leonardo Castellani}
\vskip .4cm
\centerline{\sl Istituto Nazionale di Fisica Nucleare, Sezione di Torino}
\centerline{\sl Via P. Giuria 1, I-10125 Torino, Italy}
\vskip 3cm
\centerline{\bf Abstract}
\vskip .5cm

We find two different q-generalizations of Yang-Mills theories. The
corresponding lagrangians are invariant under the q-analogue of
infinitesimal gauge transformations. We explicitly give the lagrangian
and the transformation rules for the bicovariant q-deformation of
$SU(2) \times U(1)$. The gauge potentials satisfy q-commutations, as one
expects from the differential geometry
of quantum groups. However, in one of
the two schemes we present, the field strengths do commute.

\vskip 5cm
\vbox{\hbox{DFTT-19/92}\hbox{April 1992}}

\vfill
\eject

It is tempting to try and construct gauge theories for q-deformed
Lie algebras. The motivations are manifold:
\sk
- the deformation of the classical theory could be interpreted as a
kind of symmetry breaking; however the breaking does not {\sl reduce}
the symmetry, but deforms it to a q-symmetry. This mechanism could
eventually give masses to some vector bosons without need of Higgs
fields.
\sk
- the deformation parameter $q-1$ could act as a regulator, much
as the $\epsi = D-4$ of dimensional regularization. For $q \not= 1$
the theory is invariant under the action of the quantum group:
in some sense the symmetry of the theory is ``adapted" to the
particular value $q$, changing smoothly with $q$ and becoming the
classical symmetry when $q=1$.
\sk
- in the explicit case
we will study in this letter, i.e. the q-deformation
of $SU(2) \times U(1)$, we find that the corresponding
q-Lie algebra is not a direct product any more, thus providing
a ``unification" of the electroweak group into a more general structure.
\sk
The first two points are still of speculative character. In this letter
we construct two distinct q-deformed classical theories that gauge
a given quantum group. Actually we will consider the q-analogue
of Lie algebras (bicovariant
q-Lie algebras [1], see also ref.s [2,3,4,5,6])
associated
to a given quantum group, and the variations of the gauge potentials
generalize the infinitesimal gauge transformations of the $q=1$ case.
The bicovariance conditions [1,2,5] will be essential for the consistency
of our procedure.
\sk
We stress that space-time remains ordinary (commuting) spacetime.
We do not discuss the $\hbar$-quantization of these theories.
\sk
Different approaches to the gauging of quantum groups are referred
in [7].

\sk
The first scheme we present is the most naive one: we just consider
the field strengths

$$\Fmn = \dm \An - \dn \Am + \Am \An - \An \Am  \eqno(1)$$

\noi where

$$\Am \equiv \Am^i T_i $$

\noi and the $T_i$ are the ``generators" of a bicovariant q-Lie algebra
[1]:

$$T_i T_j - \R{kl}{ij} T_k T_l = \C{ij}{k} T_k \eqno(2) $$

If we postulate the commutations

$$\Am^i \An^j= \R{ij}{kl} \An^k \Am^l  \eqno(3)$$

\noi we see that (1) can be recast in the form:

$$\Fmn^i = \dm \An^i - \dn \Am^i + \Am^j \An^k \C{jk}{i}  \eqno(4)$$

In the \q1 limit of the q-algebra (2) we have $\R{kl}{ij}=\de^k_j
\de^l_i$ and (3) just states that $\Am^i$ and $\An^j$ commute.

The next step is to define variations of the potential $\Am^i$:

$$\de \Am=-\dm \epsi - \Am \epsi + \epsi \Am \eqno(5)$$

\noi with

$$\epsi \equiv \epsi^i T_i $$

\noi and the commutations

$$\epsi^i \Am^j = \R{ij}{kl} \Am^k \epsi^l  \eqno(6)$$

Using (6) the variations become

$$\de \Am^i = - \dm \epsi^i - \Am^j \epsi^k \C{jk}{i} \eqno(7)$$

Under the variations (5) it is immediate to see that, as in
the $q=1$ case, the field strength (1) transforms as

$$\de \Fmn = \epsi \Fmn - \Fmn \epsi   \eqno(8)$$
\noi Indeed the calculation is identical to the usual one.
\sk
Now it would be nice if $\Fmn^i$ and $\epsi^j$ had the commutation
property:

$$\epsi^i \Fmn^j = \R{ij}{kl} \Fmn^k \epsi^l  \eqno(9)$$

\noi Then (8) would take the familiar form

$$ \de \Fmn^i = - \Fmn^j \epsi^k \C{jk}{i}  \eqno(10)$$

Let us prove that (9) indeed holds: by applying $\dm$ to (6) we find

$$(\dm \epsi^i) \An^j + \epsi^i \dm \An^j = \R{ij}{kl}
   [(\dm \An^k) \epsi^l + \An^k (\dm \epsi^l)] \eqno(11) $$

\noi Requesting that the $\part \epsi$ terms separately cancel
yields

$$(\dm \epsi^i)\An^j=\R{ij}{kl} \An^k (\dm \epsi^l) \eqno(12a)$$
$$\epsi^i \dm \An^j = \R{ij}{kl} (\dm \An^k)\epsi^l  \eqno(12b)$$

\noi Eq. (12b) tells us that the $\dma \Ana^i$ part of $\Fmn^i$
does satisfy (9). What happens to the quadratic piece ? We should
find that

$$\epsi^i \Am^m \An^n \C{mn}{j}=\R{ij}{kl} \Am^m \An^n \C{mn}{k}
  \epsi^l  \eqno(13)$$

\noi in order to satisfy (9). The first term
in (13) can be ordered as $AA\epsi$
by using (6) twice, and comparing it to the right hand side we find the
condition

$$\R{ir}{mk} \R{ks}{nl} \C{rs}{j}=\R{ij}{kl} \C{mn}{k} \eqno(14)$$

\noi This is one of the four conditions that characterize
bicovariant quantum
Lie algebras [1,2,5], whose quadratic relations are given in (2).
These conditions are:

$$\R{ij}{kl} \R{lm}{sp} \R{ks}{qu}=\R{jm}{kl} \R{ik}{qs} \R{sl}{up}
{\rm ~~(Yang-Baxter~equation)} \eqno(15)$$

$$\C{mi}{r} \C{rj}{n} - \R{kl}{ij} \C{mk}{r}
\C{rl}{n} = \C{ij}{k} \C{mk}{n} {\rm ~~(q-Jacobi~identities)}
\eqno(16) $$

$$\C{is}{j} \R{sq}{rl} \R{ir}{pk} + \C{rl}{q} \R{jr}{pk} = \R{jq}{ri}
\R{si}{kl} \C{ps}{r} + \R{jq}{pi} \C{kl}{i} \eqno(17)$$

$$\R{ir}{mk} \R{ks}{nl} \C{rs}{j}=\R{ij}{kl} \C{mn}{k} \eqno(18)$$

\noi The last two conditions are trivial in the limit \q1 ($\R{ij}{kl}
=\de^i_l \de^j_k$). The q-Jacobi identity shows that the matrix
$(T_i)_j^{~k} \equiv \C{ji}{k}$ is a representation (the adjoint
representation) of the q-algebra (2).
\sk
Also the third condition (17) plays an important role: it
ensures that the commutation relations (3) are preserved by the q-gauge
transformations (7). Indeed applying (7) to both sides of (3) yields

$$\de \Am^i \An^j + \Am^i \de \An^j = \R{ij}{kl} ( \de \An^k \Am^l
+ \An^k \de \Am^l)  \eqno(19)$$

\noi If we omit the  $\part \epsi$ terms, the resulting relation
can easily be seen to coincide with the bicovariance condition
(17), once all the terms have been ordered as $AA\epsi$. The
$\part \epsi$ terms partly cancel because of (12a). To cancel them
completely, we need the relation

$$\R{ij}{kl} \R{kl}{rs}= \delta^i_r \delta^j_s  \eqno(20)$$

\noi which further restricts our bicovariant q-algebra \footnote{*}
{we thank P. van Nieuwenhuizen for pointing this out.}.
\sk
A set of $\R{ij}{kl}$ and $\C{jk}{i}$
satisfying the above conditions, including (20), was found in ref. [5a].
The corresponding ``quantum Lie
algebra" is obtained by substitution of these tensors into the general
equation (2), and the result is the bicovariant q-deformation of
the D=2 Poincar\'e  Lie algebra.
This algebra may be relevant in the context of q-gravity theories
(with non-commuting space-time), but is hardly a good candidate
for interesting q-gauge theories.
\sk
In fact, we will consider next the bicovariant q-deformation of the
$SU(2) \otimes U(1)$ Lie algebra, called \qsu in the sequel
(see Table 1), satisfying conditions (15)-(18), but not (20).
The same algebra was studied in ref.s [1,4,6], in the context of
a bicovariant differential calculus on the quantum $SU(2)$ group.
Since (20) is not satisfied in this case, we take a slightly
different approach, more geometrically inspired.
The definition of field strengths is directly taken from the expression
of the q-analogue of the Cartan-Maurer equations for the bicovariant
algebra. In our \qsu example, these read (see for ex. [4]):

$$F^1 = dA^1 - q^{-3} A^+ A^-$$
$$F^2 = dA^2 + q^{-1} A^+ A^-$$
$$F^+ = dA^+ + q^{-1} A^+ (A^1 - A^2)$$
$$F^- = dA^- + q^{-1} (A^1 - A^2) A^-  \eqno(21)$$

\noi The Cartan-Maurer eqs. are just the right-hand sides of (21)
equated to zero. Here we consider the ``softening" of the quantum
group manifold, and allow the quantum curvatures to be nonvanishing
[5].

The one-forms $A^i$ have the following commutation rules [4]:

$$A^+ A^- + A^- A^+ = 0$$
$$A^1 A^+ + A^+ A^1 = 0$$
$$A^1 A^- + A^- A^1 = 0$$
$$A^2 A^+ + q^2 A^+ A^2 = (q^2 -1) A^+ A^1$$
$$A^2 A^- + q^{-2} A^- A^2 = (q^{-2} -1) A^- A^1$$
$$A^1 A^2 + A^2 A^1 = (q^{-2} -1) A^+ A^-$$
$$A^1 A^1=A^+ A^+=A^- A^- = 0$$
$$A^2 A^2 = (q^2-1) A^+ A^-  \eqno(22)$$

\noi where the products are q-wedge products[1] (see below). These
commutations are taken to hold also when $F^i \not= 0$, i.e. the
bimodule structure of the space of quantum one-forms remains
unchanged. For example, the definition of wedge product:

$$A^iA^j \equiv A^i \otimes A^j - \R{ij}{kl} A^k \otimes A^l \eqno(23)$$

\noi is rigidly kept the same even for potentials $A^i$ satisfying
(21) with $F^i \not= 0$.
\sk
By taking the exterior derivative of (21), and then using (21) and (22),
one finds the q-Bianchi identities on the two-form $F^i$:

$$dF^1 + q^{-3} F^+ A^- - q^{-3} A^+ F^- = 0 $$
$$dF^2 - q^{-1} F^+ A^- + q^{-1} A^+ F^- = 0 $$
$$dF^+ - q^{-1} F^+ (A^1 - A^2)  +q^{-1} A^+ (F^1-F^2) = 0 $$
$$dF^- - q^{-1} (F^1 - F^2)A^- + q^{-1} (A^1-A^2)F^- = 0 \eqno(24)$$

\noi These identities give us the definition of covariant
derivatives, and we
can postulate the following variations on the potentials $A^i$:

$$\de A^1 = -d\epsi^1-q^{-3} \epsi^+ A^- + q^{-3} A^+ \epsi^- $$
$$\de A^2 = -d\epsi^2+q^{-1} \epsi^+ A^- - q^{-1} A^+ \epsi^- $$
$$\de A^+ = -d\epsi^+ +q^{-1} \epsi^+ (A^1 - A^2) - q^{-1} A^+ (\epsi^1
   -\epsi^2) $$
$$\de A^- = -d\epsi^- +q^{-1} (\epsi^1 - \epsi^2) A^- - q^{-1}(A^1-A^2)
 \epsi^- \eqno(25)$$

\noi Requiring the commutations (22) to be
compatible with these variations
yields the relations between $A$ and $\epsi$ given in Table 2. Using
these
relations, we find that the
curvatures $F^i$ transform under (25) as:

$$\de F^1 = - q^{-3} \epsi^+ F^- + q^{-3} F^+ \epsi^-$$
$$\de F^2 =  q^{-1} \epsi^+ F^- - q^{-1} F^+ \epsi^-$$
$$\de F^+ = q^{-1} \epsi^+ (F^1-F^2) - \qm F^+ (\epsi^1 - \epsi^2) $$
$$\de F^- = q^{-1} (\epsi^1 - \epsi^2) F^- -
\qm(F^1-F^2)\epsi^-  \eqno(26)$$

\noi Note that in this second scheme the curvatures $F^i$
can consistently be taken to commute
with the potentials $A^i$ and with the
parameters $\epsi^j$. Indeed, as
we have stressed, the commutations between
the $A^i$ are not
modified by the ``softening" of the quantum group manifold,
i.e. by allowing $F^i \not= 0$. But then $A^i$ must commute with
the Cartan-Maurer equation, that
is with the right-hand side of eq.s (21),
which implies that it must commute with $F^j$. Similarly, the
q-commutations
of $\epsi^i$ and $A^j$ are such that $\epsi^i$ commutes with $F^j$.
\sk
Since we want to obtain a q-gauge theory on {\sl ordinary} spacetime, we
can consider our q-fields to be functions of the ordinary x-space, and

$$A^i \equiv \Am^i dx^{\mu} $$
$$F^i \equiv \Fmn^i dx^{\mu} \we dx^{\nu} \eqno(27)$$

\noi where the differentials $dx^{\mu}$ are ordinary
one-forms (whereas the
coefficients $\Am^i$ are again q-commuting objects, whose q-commutations
can be deduced from eqs. (22) by projecting on $dx^{\mu} \we dx^{\nu}$).
\sk
Finally, we construct the lagrangian invariant under the \qsu quantum
Lie algebra. We set

$$\L=\Fmn^i \Fmn^j \gij\eqno(28)$$

and determine the metric $\gij$ so that $\de \L = 0$ under
the variations (25). Here things are even simpler than in the previous
scheme since $\epsi^i$ and $F^j$ commute.

We find that:

$$\gij =
   \left( \matrix{
     2 & 2(q^{-2}-2r) & 0 & 0 \cr
     2(q^{-2}-2r)   & 2[q^{-4} + 2(1-q^{-2})r] & 0 & 0 \cr
      0             &  0 & 0 & 4r \cr
      0             &  0 & 4r & 0 \cr
   }\right)
 \eqno(29)$$

\noi solves our problem. Then $\L=\Fmn^i \Fmn^j \gij$ with
$\Fmn^i$ and $\gij$ respectively given by (21) and (29) is
invariant under
the q-gauge transformations (25).

\sk
In conclusion, we have constructed a lagrangian for the q-gauge
theory of
\qsu . We do not have an ordinary field theory, the gauge
potentials being non-commuting.
\sk
In the \q1 limit, the extra arbitrary parameter $r$ in the metric (29)
is related to the free choice of ratio of coupling constants.
The kinetic terms of the $A^1$ and $A^2$
potentials are

$$2(\dma \Ana^1 \dma \Ana^1 + \dma \Ana^2 \dma \Ana^2)+
   2(1-2r)(\dma \Ana^1 \dma \Ana^2 + \dma \Ana^2 \dma \Ana^1) \eqno(30)$$

\noi (Note that $dA^1 dA^2 = dA^2 dA^1$). In order
to diagonalize the kinetic terms we define as usual

$$A^1 = \sin \theta ~B + \cos \theta ~Z$$
$$A^2  = \cos \theta ~B - \sin \theta ~Z  \eqno(31)$$

\noi With $\theta= \pi/4$ the quadratic expression in (30)
becomes diagonal:

$$4(1-r)\dma \Bna \dma \Bna +  4r \dma \Zna \dma \Zna  \eqno(32)$$

\noi and we see that the ratio of the $U(1)$ and $SU(2)$
coupling constants
is given by

$${{\gu^2} \over {\gsu^2}}={{r}\over{1-r}} \eqno(33)  $$

\noi When $q \not= 1$, the kinetic part is

$$ 2 \dma \Ana^1 \dma \Ana^1 + [2 q^{-4} + 4(1-q^{-2})r]
\dma \Ana^2 \dma \Ana^2 +
   4(q^{-2} - 2r)\dma \Ana^1 \dma \Ana^2  \eqno(34)$$

\noi and it is straightforward to compute the angle in (31)
that diagonalizes
(34):

$$\theta={\rm arctg} [-Q \pm \sqrt{Q^2 +1}] $$
$$Q={{q^{-4}-1+2(1-q^{-2})r} \over {2(q^{-2} - 2r)}} \eqno(35)$$
\noi which can be considered the q-analogue of
the Weinberg's angle in the
electroweak theory.

The free parameter $r$ in (29) is still related to the ratio of coupling
constants:

$${{\gu^2} \over {\gsu^2}}={{2+[2q^{-4} + 4(1-q^{-2})r] (\tg \theta)^2-
4(q^{-2}-2r) \tg \theta} \over
{2+[2q^{-4} + 4(1-q^{-2})r] (\tg \theta)^{-2}+
4(q^{-2}-2r) (\tg \theta)^{-1}}} \eqno(36)$$

\noi with $\theta$ given in (35).

\sk

\vfill\eject

\centerline{{\bf Table 1}}
\sk
\centerline{The bicovariant \qsu algebra}
\sk
\sk
{\sl Non-vanishing components of the $R$ and {\bf C} tensors:}

$$\R{11}{11}=1,~~~\R{1+}{+1}=q^{-2},~~~\R{1-}{-1}=q^2,~~~\R{12}{21}=1,$$
$$\R{+1}{1+}=1,~~~\R{+1}{+1}=1-q^{-2},
{}~~~\R{++}{++}=1,~~~\R{+-}{11}=1-q^2,$$
$$\R{+-}{-+}=1,~~~\R{+-}{21}=1-q^{-2},
{}~~~\R{+2}{+1}=-1+q^{-2},~~~\R{+2}{2+}=1,$$
$$\R{-1}{1-}=1,~~~\R{-1}{-1}=1-q^{2},
{}~~~\R{-+}{11}=-1+q^2,~~~\R{-+}{+-}=1,$$
$$\R{-+}{21}=-1+q^{-2},~~~\R{--}{--}=1,
{}~~~\R{-2}{-1}=-1+q^2,~~~\R{-2}{2-}=1,$$
$$\R{21}{11}=(q^2-1)^2,~~~\R{21}{12}=1,
{}~~~\R{21}{+-}=q^2-1,~~~\R{21}{-+}=1-q^2,$$
$$\R{21}{21}=2-q^2-q^{-2},~~~\R{2+}{1+}=-q^2+q^4,~~~\R{2+}{+2}=q^2,~~~
\R{2+}{2+}=1-q^2,$$
$$\R{2-}{1-}=1-q^2,~~~
\R{2-}{-1}=-1+q^{-2}-q^2+q^4,~~~\R{2-}{-2}=q^{-2},
{}~~~\R{2-}{2-}=1-q^{-2},$$
$$\R{22}{11}=-(q^2-1)^2,~~~
\R{22}{+-}=1-q^2,~~~\R{22}{-+}=q^2-1,~~~\R{22}{21}=(q^{-1}-q)^2,$$
$$\R{22}{22}=1  $$
\sk
$$\C{11}{1}=q(q^2-1),~~~\C{11}{2}=-q(q^2-1),~~~\C{1+}{+}=q^3,~~~
  \C{1-}{-}=-q$$
$$\C{21}{1}=\qm-q,~~~\C{21}{2}=q-\qm,~~~\C{2+}{+}=-q,~~~
  \C{2-}{-}=\qm$$
$$\C{+1}{+}=-\qm,~~~\C{+2}{+}=q,~~~\C{+-}{1}=q,
{}~~~\C{+-}{2}=-q$$
$$\C{-1}{-}=q(q^2+1)-\qm,~~~\C{-2}{-}=-\qm,~~~\C{-+}{1}=-q,~~~
  \C{-+}{2}=q$$
\sk
{\sl The q-algebra:}
\sk
$$T_+ T_- - T_- T_+ + (1-q^2) T_2 T_1-(1-q^2) T_2 T_2 = q (T_1-T_2)$$
$$T_1 T_+ - T_+ T_1 -(q^4-q^2)T_2 T_+ = q^3 T_+$$
$$T_1 T_- - T_- T_1 +(q^2-1)T_2 T_- = -q T_-$$
$$T_1 T_2 - T_2 T_1 =0$$
$$T_+ T_2-q^2 T_2 T_+ = qT_+$$
$$T_- T_2-q^{-2} T_2 T_- = -\qm T_-$$
\sk

\vfill\eject
\centerline{{\bf Table 2}}
\sk
\centerline{The lagrangian of \qsu }
\sk
{\sl The lagrangian}
\sk
$$\L=\Fmn^i\Fmn^j\gij$$
\sk
{\sl The field strengths}
\sk
$$F^1 = dA^1 - q^{-3} A^+ A^-$$
$$F^2 = dA^2 + q^{-1} A^+ A^-$$
$$F^+ = dA^+ + q^{-1} A^+ (A^1 - A^2)$$
$$F^- = dA^- + q^{-1} (A^1 - A^2) A^- $$
\sk
{\sl The \qsu invariant metric}
\sk
$$\gij =
   \left( \matrix{
     2 & 2(q^{-2}-2r) & 0 & 0 \cr
     2(q^{-2}-2r)   & 2[q^{-4} + 2(1-q^{-2})r] & 0 & 0 \cr
      0             &  0 & 0 & 4r \cr
      0             &  0 & 4r & 0 \cr
   }\right)$$
\sk
{\sl The transformation rules}
\sk
$$\de A^1 = -d\epsi^1-q^{-3} \epsi^+ A^- + q^{-3} A^+ \epsi^- $$
$$\de A^2 = -d\epsi^2+q^{-1} \epsi^+ A^- - q^{-1} A^+ \epsi^- $$
$$\de A^+ = -d\epsi^+ +q^{-1} \epsi^+ (A^1 - A^2) - q^{-1} A^+ (\epsi^1
   -\epsi^2) $$
$$\de A^- = -d\epsi^- +q^{-1} (\epsi^1 - \epsi^2) A^- - q^{-1}(A^1-A^2)
 \epsi^- $$
\sk
{\sl The commutations between the 1-forms $A^i$}
\sk
$$A^+ A^- + A^- A^+ = 0$$
$$A^1 A^+ + A^+ A^1 = 0$$
$$A^1 A^- + A^- A^1 = 0$$
$$A^2 A^+ + q^2 A^+ A^2 = (q^2 -1) A^+ A^1$$
$$A^2 A^- + q^{-2} A^- A^2 = (q^{-2} -1) A^- A^1$$
$$A^1 A^2 + A^2 A^1 = (q^{-2} -1) A^+ A^-$$
$$A^1 A^1=A^+ A^+=A^- A^- = 0$$
$$A^2 A^2 = (q^2-1) A^+ A^- $$
\sk
{\sl The commutations between $\epsi^i$ and $A^j$}
\sk
$$\epsi^+ A^- - A^+ \epsi^- + \epsi^- A^+ - A^- \epsi^+ =0$$
$$\epsi^1 A^\pm - A^1 \epsi^\pm + \epsi^\pm A^1 - A^\pm \epsi^1 =0$$
$$\epsi^2 A^+ - A^2 \epsi^+ + q^2 (\epsi^+ A^2 - A^+ \epsi^2)=
  (q^2 -1) (\epsi^+ A^1 -  A^+ \epsi^1)$$
$$\epsi^2 A^- - A^2 \epsi^- + q^{-2} (\epsi^- A^2 - A^- \epsi^2)=
  (q^{-2} -1) (\epsi^- A^1 -  A^- \epsi^1)$$
$$\epsi^1 A^2 - A^1 \epsi^2 + \epsi^2 A^1 - A^2 \epsi^1=
  (q^{-2} -1) (\epsi^+ A^- -  A^+ \epsi^-)$$
$$\epsi^1 A^1 - A^1 \epsi^1=0$$
$$\epsi^+ A^+ - A^+ \epsi^+=0$$
$$\epsi^- A^- - A^- \epsi^-=0$$
$$\epsi^2 A^2 - A^2 \epsi^2=(q^2-1)(\epsi^+ A^- - A^+ \epsi^-)$$

\vfill\eject
{\bf References}
\sk
\item{[1a]} S.L. Woronowicz, Publ. RIMS, Kyoto Univ., Vol. {\bf 23}, 1
(1987) 117.
\item{[1b]} S.L. Woronowicz, Commun. Math. Phys. {\bf 111} (1987) 613;
\item{[1c]} S.L. Woronowicz, Commun. Math. Phys. {\bf 122}, (1989) 125.
\sk
\item{[2]} D. Bernard, in the notes of the E.T.H workshop, Z\"urich
1989.
\sk
\item{[3]} B. Jurco, Lett. Math. Phys. {\bf 22} (1991) 177.
\sk
\item{[4]} B. Zumino, {\sl Introduction to the Differential Geometry
of Quantum Groups}, LBL-31432 and UCB-PTH-62/91.
\sk
\item{[5a]} L. Castellani, Phys. Lett. {\bf B279} (1992) 291.
\item{[5b]} P. Aschieri and L. Castellani, {\sl An introduction to
non-commutative differential geometry on quantum groups}, Torino
preprint DFTT-20/92.
\sk
\item{[6]} W. Weich, {\sl Die Quantengruppe $SU_q (2)$ - kovariante
  Differentialrechnung und ein quantenmechanisches Modell}, Ph.D.
Thesis, Karlsruhe University, (1990).
\item{} F. M\"uller-Hoissen, {\sl Differential Calculi on the
quantum group $GL_{p,q} (2)$}, G\"ottingen preprint GOET-TP 55/91.
\sk
\item{[7]} A.P. Isaev and Z. Popowicz, {\sl q-Trace for the Quantum
Groups and q-deformed Yang-Mills Theory}, Wroclaw preprint ITP UWr
786-91.
\item{} I.Y. Aref'eva and I.V. Volovich, Mod. Phys. Lett {\bf A6}
(1991) 893; Phys. Lett. {\bf B264} (1991) 62.

\vfill\eject\end